\documentclass[aps,prd,amsmath
,eqsecnum            
,preprintnumbers
 ,nofootinbib         
 ,twocolumn         
]{revtex4}
\usepackage[dvips]{graphicx}
\usepackage{amsmath}
\usepackage{amsfonts}
\usepackage{amssymb}
\usepackage{longtable}   

\allowdisplaybreaks[4]     

\newcommand{\df}{\ {\overset {\rm def} =}\ }

\newcommand{\dril}[2]{{{\rm d} {#1}} / {{\rm d} {#2}}}

\begin{document}

\title{Exact inhomogeneous models and the drift of light rays induced by
nonsymmetric flow of the cosmic medium}

\author{Andrzej Krasi\'nski$^*$}

\affiliation{N. Copernicus Astronomical Center, Polish Academy of Sciences,\\
ul. Bartycka 18, 00 716 Warsaw, Poland\\
$^*$E-mail: akr@camk.edu.pl}

\author{Krzysztof Bolejko$^*$}

\affiliation{The Sydney Institute for Astronomy, School of Physics A28, \\
The University of Sydney, NSW 2006, Australia \\
$^*$E-mail: bolejko@physics.usyd.edu.au }

\begin{abstract}
After introducing the Szekeres and Lema\^{\i}tre--Tolman cosmological models,
the real-time cosmology program is briefly mentioned. Then, a few widespread
misconceptions about the cosmological models are pointed out and corrected.
Investigation of null geodesic equations in the Szekeres models shows that
observers in favourable positions would see galaxies drift across the sky at a
rate of up to $10^{-6}$ arc seconds per year. Such a drift would be possible to
measure using devices that are under construction; the required time of
monitoring would be $\approx10$ years. This effect is zero in the FLRW models,
so it provides a measure of inhomogeneity of the Universe. In the Szekeres
models, the condition for zero drift is zero shear. But in the shearfree normal
models, the condition for zero drift is that, in the comoving coordinates, the
time dependence of the metric completely factors out.
\end{abstract}

\maketitle

\setcounter{footnote}{1}

\section{Inhomogeneous models in astrophysics$^1$}\footnotetext{This article is based
on the talk delivered at the Marcel Grossman Meeting 13 in Stockholm,
2012.}\label{introduction}

Just as was the case with our earlier reviews \cite{Kras1997,BKHC2010,BCKr2011},
we define inhomogeneous cosmological models as those exact solutions of
Einstein's equations that contain at least a subclass of nonvacuum and nonstatic
Friedmann -- Lema\^{\i}tre -- Robertson -- Walker (FLRW) solutions as a limit.
The reason for this choice is that such FLRW models are generally considered to
be a good first approximation to a description of our real Universe, so it makes
sense to consider only those other models that have a chance to be a still
better approximation. Models that do not include an FLRW limit would not easily
fulfil this condition.

This is a live topic, with new contributions appearing frequently, so any review
intended to be complete would become obsolete rather soon. Ref. 1 is complete
until 1994, Ref. 2 is a selective update until 2009 and Ref. 3 is a more
selective update until the end of 2010. The criterion of the selection in the
updates was the usefulness of the chosen papers for solving problems of
observational cosmology.

In the very brief overview given here we put emphasis on pointing out and
correcting the erroneous results that exist in the literature and are being
taken as proven truths. Some of them have evolved to become research paradigms,
with many followers, some others proceed in this direction. We hope to stop this
process, which disturbs and slows down the recognition of the inhomogeneous
models as useful devices for understanding the observed Universe.

We first present the two classes of inhomogeneous models that, so far, proved
most fruitful in their astrophysical application: the Szekeres model
\cite{Szek1975}, and its spherically symmetric limit, the Lema\^{\i}tre
\cite{Lema1933} -- Tolman \cite{Tolm1934} (L--T) model. Then we briefly mention
the real-time cosmology program, and we give an overview of the erroneous ideas.
Finally, we present an effect newly calculated in a few classes of models that
is possible to observe and can become a test of homogeneity of the Universe: the
drift of light rays induced by nonsymmetric flow of the cosmic medium

\section{The Szekeres solution}\label{Szek}

The (quasi-spherical) Szekeres solution \cite{Szek1975,PlKr2006} is, in comoving
coordinates
\begin{equation*}
{\rm d} s^2 = {\rm d} t^2 - \frac {{\cal E}^2 {(\Phi / {\cal E}),_r}^2} {1 +
2E(r)} {\rm d} r^2 - \frac {\Phi^2} {{\cal E}^2} \left({\rm d} x^2 + {\rm d}
y^2\right),
\end{equation*}
 \vspace{-4mm}
\begin{equation}
{\cal E} \df \frac {(x - {P})^2} {2{S}} + \frac {(y - {Q})^2} {2{S}} + \frac
{{S}} 2,\label{2.1}
\end{equation}
where $E(r)$, $M(r)$, ${P}{(r)}$, ${Q}{(r)}$ and ${S}{(r)}$ are arbitrary
functions and $\Phi(t,r)$ obeys
\begin{equation}\label{2.2}
{\Phi,_t}^2 = 2E(r) + \frac {2 {M}{(r)}} {\Phi} + \frac 1 3 \Lambda \Phi^2.
\end{equation}
The source in the Einstein equations is dust, whose mass density in energy units
is
\begin{equation}\label{2.3}
\kappa \rho = \frac {2 \left(M / {\cal E}^3\right),_r} {(\Phi / {\cal E})^2
\left(\Phi / {\cal E}\right),_r}.
\end{equation}
Eq. (\ref{2.2}) implies that the bang time is in general position-dependent:
\begin{equation}\label{2.4}
\int\limits_0^{\Phi}\frac{{\rm d} \widetilde{\Phi}}{\sqrt{2E + 2M /
\widetilde{\Phi} + \frac 1 3 \Lambda \widetilde{\Phi}^2}} = t - {t_B(r)}.
\end{equation}

The general Szekeres metric has no symmetry. It contains the spherically
symmetric Lema\^{\i}tre \cite{Lema1933} -- Tolman \cite{Tolm1934} (L--T) model
as the limit of $(P, Q, S)$ being all constant. The latter is usually used with
a different parametrisation of the spheres of constant $(t, r)$, namely
\begin{equation}\label{2.5}
{\rm d}s^2 = {\rm d}t^2 - \frac{R,_r^2}{1 + 2E} {\rm d}r^2 - R^2(t,r) \left({\rm
d}\vartheta^2 + \sin^2 \vartheta {\rm d}\varphi^2 \right),
\end{equation}
where $R \equiv \Phi$, still obeying (\ref{2.2}), and (\ref{2.3}) simplifies to
\begin{equation}\label{2.6}
\kappa \rho = \frac {2 M,_r} {R^2 R,_r}.
\end{equation}

The Friedmann limit follows when, in addition, $\Phi (t,r) = r S(t)$, $2E = - k
r^2$ where $k =$ const is the FLRW curvature index, and $t_B$ is constant.

\section{Real-time cosmology}

There are ways in which the expansion of the Universe might be directly
observed. The authors of Refs. \cite{QQAm2009,QABC2012} composed them into a
paradigm termed {\bf \em real-time cosmology}. One of them is the {\bf \em
redshift drift}: the change of redshift with time for a fixed light source,
induced by the expansion of the Universe.

Consider, as an example, an L--T model with $\Lambda$, given by (\ref{2.5}) with
$R$ obeying (\ref{2.2}). Along a single radial null geodesic, directed toward
the observer, $t = T(t_o,r)$ (where $t_o$ is the instant of observation) the
redshift is \cite{Bond1947}
\begin{equation}\label{3.1}
1 + z(t_o,r) = {\rm exp} \left[\int_{r_{\rm em}}^{r_{\rm obs}} \frac
{R,_{tr}(T(t_o,r), r)} {\sqrt{1 + 2E(r)}} {\rm d} r\right].
\end{equation}

For any fixed source (i.e. constant $r$), eq. (\ref{2.2}) defines a different
expansion velocity $R,_t$ for $\Lambda = 0$ and for $\Lambda \neq 0$, and thus
allows us to calculate the contribution of $\Lambda$ to $z$ via (\ref{3.1}).
With the evolution type known, the expansion velocity depends on $r$, and so
allows us to infer the distribution of mass along the past light cone.

According to the authors of \cite{QQAm2009,QABC2012}, the ``European Extremely
Large Telescope'' (E-ELT)\footnote{Now in the planning, to be built in Chile.}
could detect the redshift drift during less than 10 years of monitoring a given
light source. The Gaia
observatory\footnote{http://sci.esa.int/science-e/www/area/index.cfm?fareaid=26}
could achieve this during about 30 years.

For more on redshift drift see the contribution by P. Mishra, M.-N. C\'el\'erier
and T. Singh in these Proceedings.

The drift of light rays described in Sec. \ref{Szredshift} and following is
another real-time cosmology effect.

\section{Erroneous ideas and paradigms}\label{errideas}

The L--T model was noticed in the astrophysics community -- but:

1. Many astrophysicists treat it as an enemy to kill rather than as a useful new
device. (Citation from Ref. \cite{QABC2012}: The Gaia or E-ELT projects could
distinguish FLRW from L--T ``possibly eliminating an {\em exotic alternative
explanation to dark energy}'').

2. Some astrophysicists practise a loose approach to mathematics. An extreme
example is to take for granted every equation found in any paper, without
attention being paid to the assumptions under which it was derived.

Papers written in such a style planted errors in the literature, which then came
to be taken as established facts. In this section a few characteristic errors
are presented (marked by {\Huge \bf {$\bullet$}}) together with their
explanations (marked by {\Huge \bf {$*$}}).

\bigskip

\noindent {\Huge \bf {$\bullet$}} \hspace {3mm} The accelerating expansion of
the Universe is an observationally established fact (many refs., the Nobel
Committee among them).

\noindent {\Huge \bf {$*$}} \hspace {3mm} The established fact is the {\em
smaller than expected observed luminosity of the SNIa supernovae} (but even this
is obtained assuming that FLRW is the right cosmological model). {\bf \em The
accelerating expansion is an element of theoretical explanation of this
observation.} When the SNIa observations are interpreted against the background
of a suitably adjusted L--T model, they can be explained by matter
inhomogeneities along the line of sight, with decelerating expansion
\cite{KHBC2010,INNa2002}.

\bigskip

\noindent {\Huge \bf {$\bullet$}} \hspace {3mm} Positive sign of the redshift
drift is a direct confirmation of accelerated expansion of the space.

\noindent {\Huge \bf {$*$}} \hspace {3mm} The sign of redshift drift is only
related to acceleration when homogeneity is assumed\cite{YKN2011} (see also
Mishra, C\'el\'erier, and Singh in these Proceedings).

\bigskip

\noindent {\Huge \bf {$\bullet$}} \hspace {3mm} $H_0$, supernovae and the cosmic
microwave background radiation (i.e. the size of the sound horizon and the
location of the acoustic peaks) are sufficient to rule out inhomogeneous L--T
models.

\noindent {\Huge \bf {$*$}} \hspace {3mm} These observations only depend on
$D(z)$ and $\rho(z)$, and thus can be accommodated by the L--T model, which is
specified by 2 arbitrary functions. Examples of such constructions are given in
\cite{CBKr2010,INNa2002}. In fact, as follows from the Sachs equations, in the
approximation of small null shear (which for most L--T models works quite
well\cite{BoFe2012}), there is a relation between $D(z)$, $\rho(z)$ and $H(z)$,
meaning that these 3 observables are not independent\cite{BoFe2012}, and thus
allow the L--T model to accommodate more data on the past null cone.

\bigskip

\noindent {\Huge \bf {$\bullet$}} \hspace {3mm} The gravitational potential of a
typical structure in the Universe is of the order of $10^{-5}$ and thus, by
writing the metric in the conformal Newtonian gauge, one immediately shows that
inhomogeneities can only introduce minute ($\approx 10^{-5}$) deviations from
the RW geometry. This also means that the evolution of the Universe must be
Friedmannian (common argument among cosmologists).

\noindent {\Huge \bf {$*$}} \hspace {3mm} Even if the gravitational potential
remains small, its spatial derivatives do not, and thus the model has completely
different optical properties than the Friedmann models. This was shown by
rewriting one of the L--T Gpc-scale inhomogeneous models in the conformal
Newtonian coordinates \cite{EMR2009}. The gravitational potential of this model
remains small yet the distance--redshift relation deviates strongly from that
for the background model.

The question remains whether small-scale fluctuations (of the order of tens of
Mpc) could also modify optical properties and evolution of the Universe. The
problem is complicated as it requires solving the Einstein equations and null
geodesics for a general matter distribution, which, with current technology, is
not possible to do numerically. Therefore, the problem has been addressed in a
number of approximations and toy models. Recent studies showed that the optical
properties along a single line of sight can be significantly different than in
the Friedmann model. Yet, if averaged over all directions, the average
distance--redshift relation closely follows that of the model, which describes
the evolution of the average density and expansion rate\cite{BoFe2012}. Thus,
the problem reduces to the following one: do the average density and expansion
rate follow the evolution of the homogeneous model, i.e. is the evolution of the
background affected by small-scale inhomogeneities and does it deviate from the
Friedmannian evolution? Some authors claim that matter inhomogeneities cannot
affect the background and that the Universe must have Friedmannian
properties\cite{IsWa2006,GrWa2011}, while others argue for strong deviation from
the Friedmannian evolution\cite{Wilt2011,RBCO2011}. Studies of this problem
within the exact models, like L--T, proved that under certain conditions the
back-reaction can be large, while under others it remains quite small
\cite{Suss2011}, leaving the problem unsolved. For an informative description of
the problem and techniques used to address it see Ref. \cite{BuRa2012,CELU2011}.

\bigskip

\noindent {\Huge \bf {$\bullet$}} \hspace {3mm} Fitting an L--T model to number
counts or the $D_L(z)$ relation results in predicting a huge void, several
hundred Mpc in radius, around the centre (too many papers to be cited,
literature still growing). Measurements of the dipole component of the CMB
radiation then imply that our Galaxy should be very close to the center of this
void, which contradicts the ``cosmological principle''.

\noindent {\Huge \bf {$*$}} \hspace {3mm} {\bf \em The implied huge void is a
consequence of handpicked constraints imposed on the L--T model}, for example
constant $t_B$. When the model is employed at full generality, the giant void is
not implied \cite{CBKr2010}.

\noindent {\Huge \bf {$*$}} \hspace {3mm} {\bf \em The cosmological principle is
a postulate, not a law of Nature}, it cannot say which model is ``right'' and
which is ``wrong''.

\bigskip

\noindent {\Huge \bf {$\bullet$}} \hspace {3mm} The bang time function must be
constant, otherwise the decaying mode of density perturbation is nonzero, which
implies large inhomogeneities in the early universe (see, for example, Ref.
\cite{ZMSc2008}).

\noindent {\Huge \bf {$*$}} \hspace {3mm} The bang time function describes the
differences in the age between different regions of the Universe. In the L--T
and Szekeres models it is also related to the amplitude of the decaying mode.
However, these models describe the evolution of dust and therefore cannot be
extended to times before the recombination, when the Universe was in a turbulent
state: rotation, plasma, pressure gradients all did affect the proper time of an
observer (${\rm d} \tau = {\rm d} t~ \sqrt{g_{00} (t,x^i)}$). Eventually, even
if the Universe started with a simultaneous big bang, by the time of
recombination, due to the standard physical processes, the age of the Universe
would have been different at different spatial positions, giving rise to
non-constant $t_B(r)$ of the dust L--T or Szekeres model that takes over there.

Moreover, the relation between the nonsimultaneous big bang and the decaying
mode was established only for the L--T \cite{Silk1977,PlKr2006} and Szekeres
\cite{GoWa1982} models. For more general models, not yet explicitly known as
solutions of Einstein's equations, like the ones mentioned above, the connection
may be more complicated and indirect. Thus, citing this relation for such a
general situation is an illegitimate stretching of a theorem beyond the domain
of its assumptions (see also the next entry below).

\bigskip

\noindent {\Huge \bf {$\bullet$}} \hspace {3mm} The L--T models used to explain
away dark energy must have their bang-time function constant, or else they ``can
be ruled out on the basis of the expected cosmic microwave background spectral
distortion'' \cite{Zibi2011}.

\noindent {\Huge \bf {$*$}} \hspace {3mm} {\bf \em The papers that claim this
parametrise their models with a set of very simple functions, which lack
flexibility}. This sets them on the wrong track from the beginning.

\noindent {\Huge \bf {$*$}} \hspace {3mm} {\bf \em In order to meaningfully test
any cosmological model against observations}, {\bf \em one must apply it at
every step of analysis} of the observational data. To do so, would require a
re-analysis of a huge pool of data. C. Hellaby with coworkers \cite{McHe2008} is
working on such a program applied to the L--T model, but the work is far from
being completed.

Lacking any better chance, we currently use observations interpreted in the FLRW
framework to infer about the $M(r)$ and $t_B(r)$ functions in the L--T model.
This is justified as long as we intend to point out possibilities, under the
tacit assumption that these results will be verified in the future within a
complete revision of the observational material on the basis of the L--T model.
However, putting ``precise'' bounds on the L--T model functions using the
self-inconsistent mixture of FLRW/L--T data available today is a self-delusion.
An example: the spatial distribution of galaxies and voids is inferred from the
luminosity distance vs. redshift relation that applies {\em only} in the FLRW
models. Without assuming the FLRW background, we know nothing about this
distribution until we reconstruct it using the L--T model from the beginning.

The L--T and Szekeres models cannot be treated as exact models of the Universe,
to be taken literally in all their aspects. They are {\em exact as solutions of
Einstein's equations}, but when applied in cosmology, they are merely {\em the
next step of approximation after FLRW}. If the FLRW approximation is good for
some purposes, then a more detailed model, {\em when applied in a situation, in
which its assumptions are fulfilled}, can only be better.

\section{The redshift equations in the Szekeres models}\label{Szredshift}

Consider two light rays, the second one following the first after a short
time-interval $\tau$, both emitted by the same source and arriving at the same
observer. The trajectory of the first ray is given by
\begin{equation}\label{5.1}
(t, x, y) = (T(r), X(r), Y(r)),
\end{equation}
the corresponding equation for the second ray is
\begin{equation}\label{5.2}
(t, x, y) = (T(r) + \tau(r), X(r) + \zeta(r), Y(r) + \psi(r)).
\end{equation}
This means that while the first ray intersects a hypersurface $r = r_0$ at $(t,
x, y) = (T, X, Y)$, the second ray intersects the same hypersurface not only
later, but, in general, at a different comoving location. \noindent {\bf \em
$\Longrightarrow$ In general the two rays will intersect different sequences of
intermediate matter worldlines}.

The same is true for nonradial rays in the L--T model.

Consequently, the second ray is emitted in a different direction and is received
from a different direction by the observer. Thus, a typical observer in a
Szekeres spacetime should see each light source slowly {\bf \em drift across the
sky}. How slowly will be estimated further on. As will be seen from the
following, {\bf \em the absence of this drift is a property of exceptionally
simple geometries} (or exceptional directions in more general geometries).

We assume that $(\zeta, \psi)$ and $(\dril {} r) (\tau, \zeta, \psi)$ are small
of the same order as $\tau$, so we neglect all terms nonlinear in any of them
and terms involving their products.

For any function $f(t, r, x, y)$ the symbol $\Delta f$ will denote
\begin{equation}\label{5.3}
f(t + \tau, r, x + \zeta, y + \psi) - f(t, r, x, y)
\end{equation}
{\bf \em linearized in $(\tau, \zeta, \psi)$.} Note: {\bf \em the difference is
taken at the same value of $r$}. Applying $\Delta$ to the null geodesic
equations parametrised by $r$ we obtain the equations of propagation of $(\tau,
\zeta, \psi)$ and $(\xi, \eta) \df (\dril {} r)$ $(\zeta, \psi)$ along a null
geodesic -- see both sets of equations fully displayed in Ref. \cite{KrBo2011}.

\section{Repeatable light paths}\label{repeat}

There will be no drift when, for a given source--observer pair, each light ray
will proceed through the same intermediate sequence of matter world lines. Rays
having this property will be called {\bf \em repeatable light paths (RLP)}. For
a RLP we have
\begin{equation}\label{6.1}
\zeta = \psi = \xi = \eta = 0
\end{equation}
all along the ray. The equations of propagation of $(\tau, \zeta, \psi, \xi,
\eta)$ become then overdetermined (3 equations to determine the propagation of
$\tau$ along a null geodesic), and imply limitations on the metric components.
They can be used in 2 ways:

1. As the condition (on the metric) for {\bf \em all} null geodesics to be RLPs.

2. As the conditions under which special null geodesics are RLPs in subcases of
the Szekeres spacetime.

In the first interpretation, the equations of propagation should be identities
in the components of $\dril {x^{\alpha}} r$, and this happens when
\begin{equation}\label{6.2}
\Psi \df \Phi,_{tr} - \Phi,_t \Phi,_r/\Phi = 0.
\end{equation}
This means zero shear, i.e. the Friedmann limit. Thus, we have the following

{\bf Corollary:}

{\bf \em The only spacetimes in the Szekeres family in which {all} null
geodesics have repeatable paths are the Friedmann models.}

In fact, we proved something stronger. Since the drift vanishes in the Friedmann
models, we have:

{\bf Corollary 2:}

{\bf \em The presence of the drift would be an observational evidence for the
Universe to be inhomogeneous on large scales.}

In the second interpretation, there are only 2 nontrivial (i.e.
non-Friedmannian) cases:

A. When the Szekeres spacetime is axially symmetric ($P$ and $Q$ are constant).
In this case, the RLPs are those null geodesics that stay on the axis of
symmetry in each 3-space of constant $t$.

B.  When the Szekeres spacetime is spherically symmetric ($P, Q, S$ are all
constant) -- then it reduces to the L--T model. In this case, the radial null
geodesics are {\em the only} RLPs that exist. A formal proof of this statement
is highly complicated \cite{KrBo2011}.

For non-radial rays in the L--T model, the non-RLP phenomenon was predicted, by
a different method (and under the name of {\bf \em cosmic parallax}), by
Quercellini {\it et al.} \cite{QQAm2009,QABC2012}. Their review contains a broad
presentation of the real-time cosmology paradigm, oriented toward observational
possibilities.

\vspace{-8mm}

\begin{center}
\begin{figure*}
\includegraphics{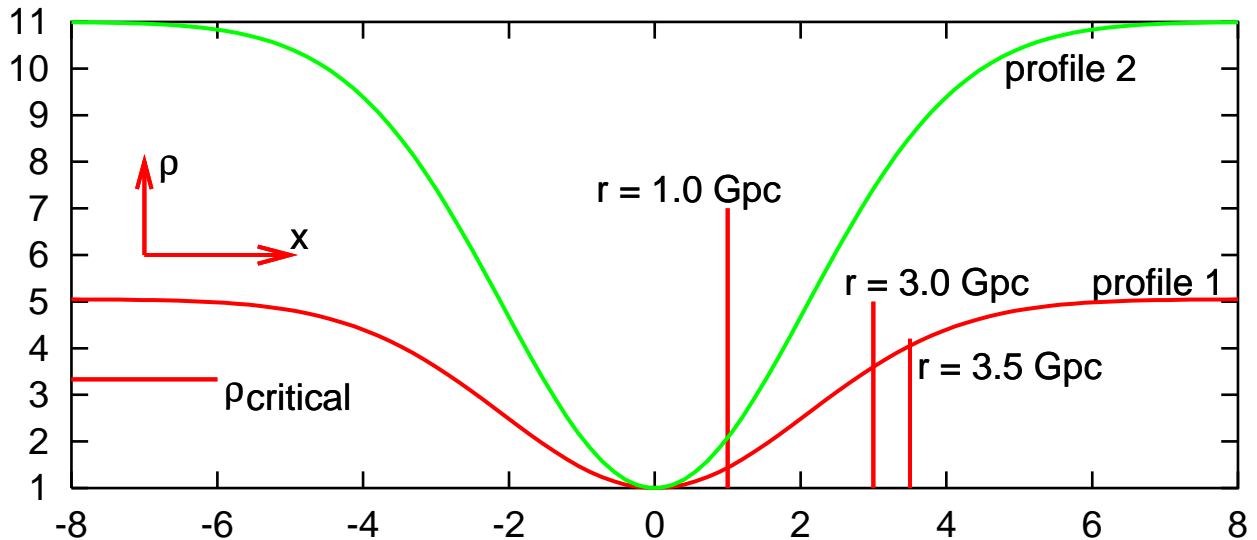} \caption{Density profiles and
positions of the observers used in the numerical
examples.}\label{densprof}
\end{figure*}
\end{center}

\section{Examples of non-RLPs in the L--T model}\label{numex}

\begin{figure}[h]
\begin{center}
\vspace{-4cm}
\includegraphics[scale=0.5]{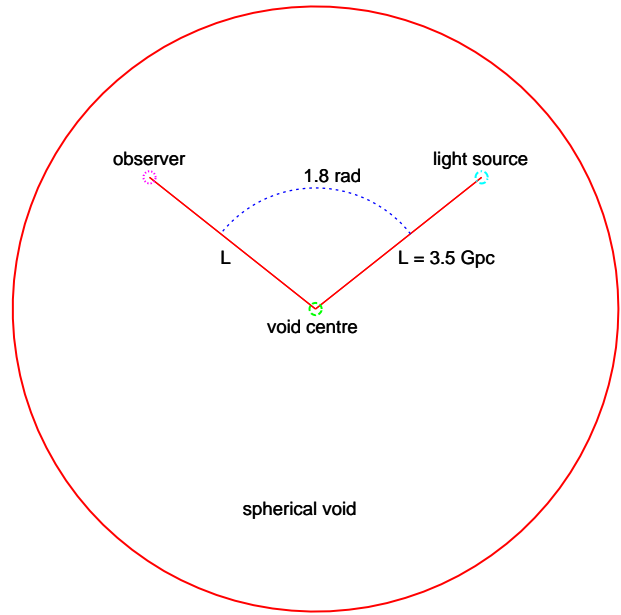}
\caption{The configuration of the light source and the observer in example
1.}\label{example1}
\end{center}
\vspace{-0.5cm}
\end{figure}

The examples will show the non-RLP effect for nonradial null geodesics in two
configurations of the L--T model, shown in Fig. \ref{densprof}, for different
positions of the observers with respect to the center of symmetry. In Example 1
(see Fig. \ref{example1}) we use Profile 1, the observer and the light source at
3.5 Gpc from the center of the void, the directions to them at the angle 1.8
rad, and \cite{BoWy2008}

$t_B =0$ -- simultaneous Big Bang,

$\rho(t_0,r) = \rho_0 \left[ 1 + \delta - \delta \exp \left( - {r^2}/{\sigma^2}
\right) \right]$ -- the density profile at the current instant,

$r \df R(t_0,r)$ -- the radial coordinate,

$\rho_0 \df \rho(t_0,0) = 0.3 \times (3H_0^2)/(8 \pi G) \equiv 0.3 \rho_{\rm
critical}$ -- the present density at the center of the void,

$H_0 = 72$~km~s$^{-1}$~Mpc$^{-1}$ -- the present value of the Hubble parameter,

$\delta = 4.05$, $\sigma = 2.96$ Gpc.

\begin{figure*}
\begin{center}
\includegraphics{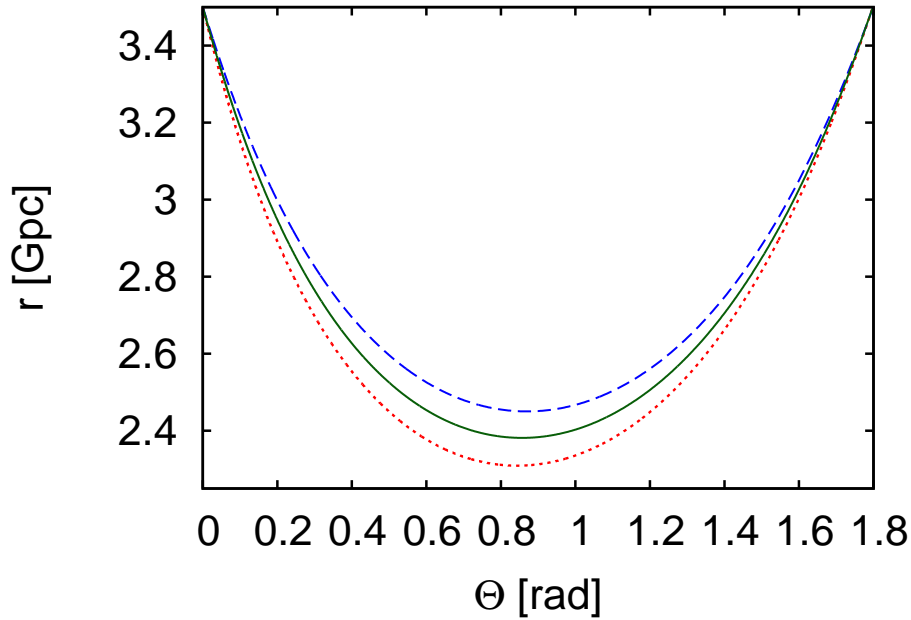} \caption{Three light rays projected on the space of
constant comoving time along the flow lines of the cosmic medium. Middle line:
the ray received at the current instant; upper line: the ray received $5 \times
10^9$ years ago; lower line: the ray to be received $5 \times 10^9$ years in the
future.}\label{fig1}
\end{center}
\end{figure*}

The source in Fig. \ref{example1} sends three light rays to the same observer,
the first of which was received by the observer $5 \times 10^9$ years ago, the
second is being received right now, and the third one will be received $5 \times
10^9$ years in the future. Fig. \ref{fig1} shows these rays projected on the
space $t =$ now along the flow lines of the L--T dust. The source is at the
upper right corner of the graph and the observer is at the upper left corner.

The {\bf \em time-averaged} rate of change of the position of the source in the
sky, seen by the observer is
\begin{eqnarray}\label{7.1}
&& \dot{\gamma} = \frac {\rm angle\ between\ the\ earlier\ and\ the\ later\ ray}
{\rm time\ interval = 5 \times 10^9\ years} \nonumber \\
&& \sim 10^{-7}\ \frac {\rm arc sec} {\rm year}.
\end{eqnarray}
The rate of drift in the next figures is calculated in the same way.

\begin{figure}[h]
\begin{center}
\includegraphics[scale=0.7]{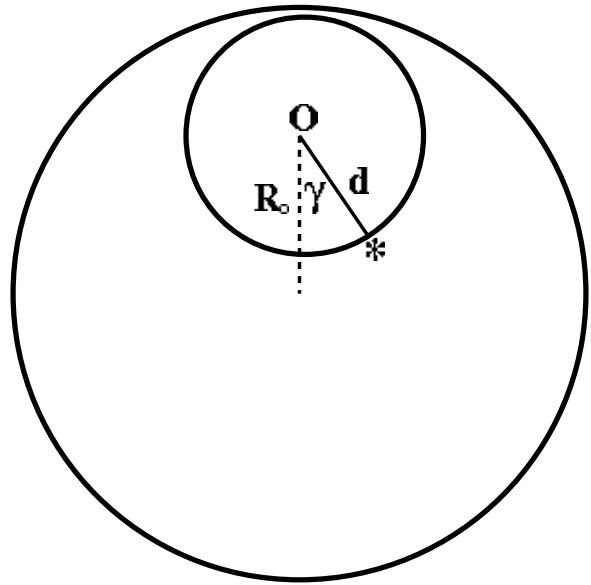} \caption{The configuration for Examples 2, 3,
4.}\label{fig2}
\end{center}
\end{figure}

\begin{figure*}
\begin{center}
\includegraphics{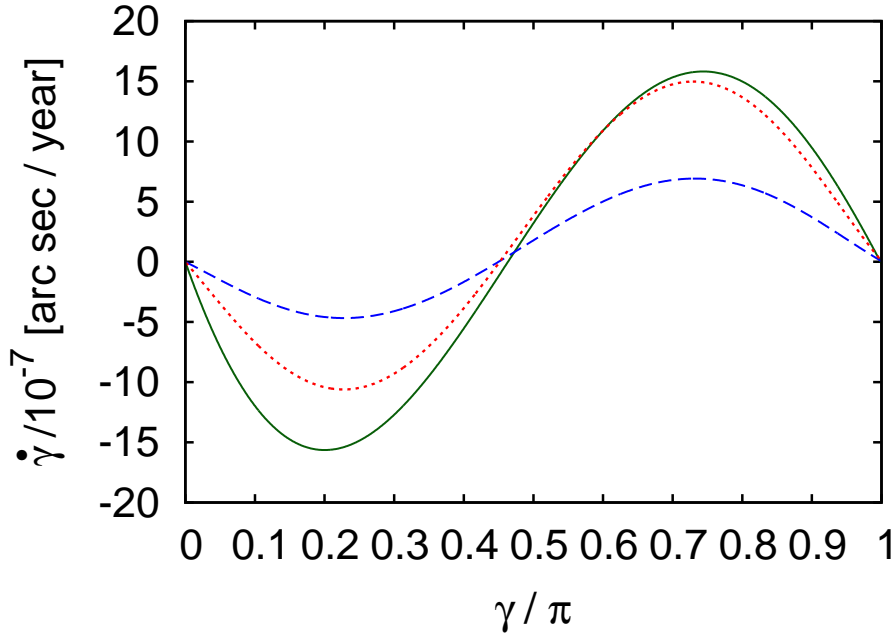} \caption{$\dot{\gamma}$ as a function of
$\gamma$ for Examples 2, 3, 4, in arcsec/(year $\times 10^{7})$.}\label{fig3}
\end{center}
\end{figure*}

In Examples 2, 3 and 4 (Fig. \ref{fig2}), the observer O is at $R_0$ from the
center; the angle between the direction toward the galaxy ({\Huge $_*$}) and
toward the origin is $\gamma$. For each $\gamma$ we calculated the rate of
change $\dot \gamma$ by eq. (\ref{7.1}), and the graphs in Fig. \ref{fig3} show
$\dot{\gamma}$ as a function of $\gamma$. All the examples have $d = 1$ Gly
$\approx 306.6$ Mpc. Example 2 (solid line) has $R_0 = 3$ Gpc and Profile 1;
Example 3 (dashed line) has $R_0 = 1$ Gpc and Profile 1; Example 4 (dotted line)
has $R_0 = 1$ Gpc and Profile 2 (for which $\delta = 10.0$ instead of $\delta =
4.05$; this is a deeper void in a higher-density background). The amplitude is
$\sim 10^{-7}$ for (3) and $\sim 10^{-6}$ for (2) and (4). With the Gaia
accuracy of $5-20 \times 10^{-6}$ arcsec, we would need a few years to detect
this effect.

\section{RLPs in shearfree normal models}\label{BarnesRLP}

In the Szekeres models, the condition for all null geodesics to be RLPs was the
vanishing of shear. This suggests that the cause of the non-RLP phenomenon might
be shear in the cosmic flow. To test this supposition, the existence of RLPs was
investigated in those cosmological models in which shear is zero \cite{Kras2011}
-- the shearfree normal models found by Barnes \cite{Barn1973}. They obey the
Einstein equations with a perfect fluid source and contain, as the
acceleration-free limit, the whole FLRW family.

There are four classes of them: the Petrov type D metrics that are spherically,
plane and hyperbolically symmetric, and the conformally flat metric found
earlier by Stephani \cite{Step1967}. In the Petrov type D case, the metric in
comoving coordinates is
\begin{equation}
{\rm d} s^2 = \left(\frac {F V,_t} V\right)^2 {\rm d} t^2 - \frac 1 {V^2}
\left({\rm d} x^2 + {\rm d} y^2 + {\rm d} z^2\right), \label{8.1}
\end{equation}
where $F(t)$ is an arbitrary function, related to the expansion scalar $\theta$
by $\theta = 3 / F$. The Einstein equations reduce to the single equation:
\begin{equation}
w,_{uu} /w^{2} = f(u), \label{8.2}
\end{equation}
where $f(u)$ is an arbitrary function, while $u$ and $w$ are related to $(x, y,
z)$ and to $V(t,x,y,z)$ differently in each subfamily. We have
\begin{equation}
(u, w) = \left\{ \begin{array}{ll} (r^{2}, V)
   & \mbox{with spherical symmetry}, \\
 \  & \ r^2 \df x^2 + y^2 + z^2;\\
(z, V)  & \mbox{with plane symmetry};\\
(x/y, V/y)  & \mbox{with hyperbolic symmetry}.\end{array} \right. \label{8.3}
\end{equation}
The FLRW limit follows when $f = 0$ and $V = R(t) g(x,y,z)$.

The conformally flat Stephani solution \cite{Step1967,Kras1997} has the metric
given by (\ref{8.1}), the coordinates are comoving, and $V(t, x, y, z)$ is given
by
\begin{eqnarray}
&&V = \frac 1 R \left\{1 + \frac 1 4 k(t) \left[\left(x - x_{0}(t)\right)^{2} +
\left(y - y_{0}(t)\right)^{2} \right.\right. \nonumber \\
&&\ \ \ \ \ \ \ \ \ \ \ \ \ \ \ \ \ \ \  + \left.\left.\left(z -
z_{0}(t)\right)^{2}\right]\right\},
\label{8.4}
\end{eqnarray}
where $(R, k, x_0, y_0, z_0)$ are arbitrary functions of $t$. This a
generalisation of the whole FLRW class, which results when $(k, x_0, y_0, z_0)$
are all constant. In general, (\ref{8.4}) has no symmetry.

In these models, in the most general cases, generic null geodesics are not RLPs.
Consequently, {\bf \em it is not shear that causes the non-RLP
property}.\footnote{Contrary to what the authors of Ref. \cite{QABC2012} claim
throughout their paper.}

In the general type D shearfree normal models, the only RLPs are radial null
geodesics in the spherical case and their analogues in the other two cases. In
the most general Stephani spacetime, RLPs do not exist. In the axially symmetric
subcase of the Stephani solution the RLPs are those geodesics that intersect the
axis of symmetry in every space of constant time. In the spherically-, plane-
and hyperbolically symmetric subcases, the RLPs are the radial geodesics.

The completely drift-free subcases are conformally flat, but more general than
FLRW. Their defining property is that their time-dependence in the comoving
coordinates can be factored out, and the cofactor metric is static. The FLRW
models have the same property. For example, in the drift-free spherically
symmetric type D case:
\begin{eqnarray}
{\rm d} s^2 = \frac 1 {V^2} && \hspace{-3mm} \left\{\left[\left(A_1 + A_2
r^2\right) \left(F S,_t {\rm d} t\right)\right]^2 - {\rm d} r^2\right.
\nonumber \\
&&- \left.r^2 \left({\rm d} \vartheta^2 + r^2
\sin^2 \vartheta {\rm d} \varphi^2\right)\right\}, \label{8.5}
\end{eqnarray}
the whole non-staticity is contained in $V$:
\begin{equation}\label{8.6}
V = B_1 + B_2 r^2 + \left(A_1 + A_2 r^2\right) S(t).
\end{equation}
The $(A_1, A_2, B_1, B_2)$ are arbitrary constants and $S(t)$ is an arbitrary
function. This model is more general than FLRW because the pressure in it is
spatially inhomogeneous. The FLRW limit follows when $A_1 \neq 0$ and $B_2 =
(A_2/A_1) B_1$.

\section{Dependence of RLPs on the observer congruence}

The RLPs are defined relative to the congruence of worldlines of the observers
and light sources. So far, we have considered observers and light sources
attached to the particles of the cosmic medium, whose velocity field is defined
by the spacetime geometry via the Einstein equations. But we could as well
consider other timelike congruences, or spacetimes in which no preferred
timelike congruence exists, for example Minkowski. It turns out that even in the
Minkowski spacetime one can devise a timelike congruence that will display the
non-RLP property \cite{Kras2012}.

Take the Minkowski metric in the spherical coordinates
\begin{equation}\label{9.1}
{\rm d} s^2 = {\rm d} {t'}^2 - {\rm d} {r'}^2 - {r'}^2 \left({\rm d} \vartheta^2
+ \sin^2 \vartheta {\rm d} \varphi^2\right),
\end{equation}
and carry out the following transformation on it:
\begin{equation}\label{9.2}
t' = (r - t)^2 + 1 / (r + t)^2, \qquad r' = (r - t)^2 - 1 / (r + t)^2.
\end{equation}
The result is the metric
\begin{eqnarray}
&& {\rm d} s^2 = \frac 1 {(r + t)^4}\left\{16 u \left({\rm d} t^2 - {\rm d}
r^2\right)\right. \\
&& \ \  \left.- \left(u^2 - 1\right)^2 \left({\rm d} \vartheta^2 + \sin^2
\vartheta {\rm d} \varphi^2\right)\right\}, \qquad u \df r^2 - t^2. \nonumber
\label{9.3}
\end{eqnarray}
Now we assume that the curves with the unit tangent vector field $u^{\alpha} =
\left[(r + t)^2 / \left(4 \sqrt{u}\right)\right] {\delta^{\alpha}}_0$ are world
lines of test observers and test light sources.

Proceeding as before we conclude that, with respect to this congruence, generic
null geodesics in the Minkowski spacetime have non-repeatable paths. (The
exception are those rays that are radial in the coordinates of (\ref{9.3})).
This is because the time-dependence of (\ref{9.3}) cannot be factored out.

\section*{Acknowledgments}

We are grateful to Marie No\"{e}lle C\'el\'erier for several helpful comments.

\end{document}